\documentclass[aps,prl,twocolumn,groupedaddress,showpacs]{revtex4}
\usepackage{graphicx}

\begin{document}

%\preprint{}

\title{Quantum and classical mode softening near the charge-density-wave/superconductor transition of Cu$_{x}$TiSe$_{2}$: Raman spectroscopic studies}



\author{H. Barath,$^1$ M. Kim,$^1$ J.F. Karpus,$^1$ S.L. Cooper,$^1$ P. Abbamonte,$^1$ E. Fradkin,$^1$ E. Morosan,$^2$ and R.J. Cava$^2$}

%\email[Electronic address: ]{karpus@uiuc.edu}

%\thanks{}
\affiliation{
$^{1}$Department of Physics and Frederick Seitz Materials Research Laboratory, University of Illinois, Urbana, Illinois  61801, USA\\
$^{2}$Department of Chemistry, Princeton University, Princeton, New Jersey  08544, USA\\
}

\date{\today}

\begin{abstract}

Temperature- and x-dependent Raman scattering studies of the charge
density wave (CDW) amplitude modes in Cu$_{x}$TiSe$_{2}$ show that
the amplitude mode frequency $\omega_{o}$ exhibits identical
power-law scaling with the reduced temperature, T/T$_{\rm CDW}$, and
the reduced Cu content, x/x$_{c}$, i.e., $\omega_{o}$ $\sim$ (1 -
p)$^{0.15}$ for p = T/T$_{\rm CDW}$ or x/x$_{c}$, suggesting that
mode softening is independent of the control parameter used to
approach the CDW transition.  We provide evidence that x-dependent
mode softening in Cu$_{x}$TiSe$_{2}$ is caused by the reduction of
the electron-phonon coupling constant $\lambda$ due to expansion of
the lattice, and that x-dependent `quantum' (T $\sim$ 0) mode
softening reveals a quantum critical point within the superconductor
phase of Cu$_{x}$TiSe$_{2}$.

\end{abstract}

\pacs{71.45.Lr, 73.43.Nq, 74.70.-b, 78.30.-j}

\maketitle


One of the most important current goals of condensed matter physics research involves elucidating the competition between diverse and exotic phases in strongly correlated matter, such as antiferromagnetism and superconductivity (SC) in the high T$_{c}$ cuprates,\cite{1} heavy fermions,\cite{2} and cobaltates,\cite{3} and charge density wave (CDW) order and SC in materials such as Na$_{x}$TaS$_{2}$.\cite{4}  Recently, Morosan \emph{et al.} discovered an interesting new material exhibiting a competition between CDW order and SC:  copper intercalated 1$T$-TiSe$_{2}$, i.e., Cu$_{x}$TiSe$_{2}$.\cite{5}  1$T$-TiSe$_{2}$ is a semimetal or small-gap semiconductor in the normal state,\cite{6,7,8,9} which develops a commensurate CDW with a 2a$_{o}$$\times$2a$_{o}$$\times$2c$_{o}$ superlattice structure at temperatures below a second-order phase transition at T$_{\rm CDW}$ $\sim$ 200 K.\cite{6,10}  Increasing Cu intercalation in TiSe$_{2}$ (increasing x in Cu$_{x}$TiSe$_{2}$) results in (i) an expansion of the a- and c-axis lattice parameters,\cite{5} (ii) increased electronic density of states near the L point,\cite{7,8} (iii) a suppression of the CDW transition temperature,\cite{5} and (iv) the emergence near x = 0.04 of a SC phase having a maximum T$_{c}$ of 4.15 K at x = 0.08.\cite{5}

The Cu$_{x}$TiSe$_{2}$ system provides an ideal opportunity to
investigate the microscopic details of quantum (T $\sim$ 0) phase
transitions between CDW order and SC.  It is of particular interest
to clarify the nature of the ``soft mode'' in CDW/SC transitions:
the behavior of the soft mode - i.e., the phonon mode whose
eigenvector mimics the CDW lattice distortion, and hence whose
frequency tends towards zero at the second-order phase transition -
is one of the most fundamental and well-studied phenomena associated
with classical (thermally driven) displacive phase
transitions;\cite{11} on the other hand, soft mode behavior
associated with quantum phase transitions is not well understood. In
this investigation, we use Raman scattering to study the
temperature- and doping-dependent evolution of the CDW `amplitude'
modes in Cu$_{x}$TiSe$_{2}$.  The CDW amplitude mode \cite{12} -
which is associated with collective transverse fluctuations of the
CDW order parameter - offers detailed information regarding the
evolution and stability of the CDW state and the CDW soft mode.  In
this study, we show that the amplitude mode frequency in
Cu$_{x}$TiSe$_{2}$ exhibits identical power-law scaling with the
reduced temperature, T/T$_{CDW}$, and the reduced Cu content,
x/x$_{c}$, indicating that mode softening in Cu$_{x}$TiSe$_{2}$ is
independent of the control parameter used to approach the CDW
transition. Further, we show that `quantum' (T $\sim$ 0) softening
of the CDW amplitude mode is consistent with a quantum critical
point hidden in the superconductor phase of Cu$_{x}$TiSe$_{2}$,
suggesting a possible connection between quantum criticality and
superconductivity.

Raman scattering measurements were performed on high quality single-crystal
and pressed-pellet samples of Cu$_{x}$TiSe$_{2}$ for x = 0, 0.01, 0.02, 0.03, 0.04, 0.05, and 0.06, which were grown and characterized as described previously.\cite{5,13}  Fig. 1(a) shows the T = 6 K Raman spectra of Cu$_{x}$TiSe$_{2}$ for various Cu concentrations (x).  The T = 6 K Raman spectrum of TiSe$_{2}$ (top spectrum in Fig. 1) exhibits several spectroscopic features that have been reported previously,\cite{14,15} including a Raman-active {\bf k} = 0 phonon mode near 137 cm$^{-1}$ that shows little change in energy ($\sim$ 0.7\%), and only a slight increase in linewidth, with increasing x.  Also apparent in Fig. 1(a) are several modes that appear below the CDW transition, including an A$_{1g}$-symmetry amplitude mode near 118 cm$^{-1}$, which arises from fluctuations of the CDW amplitude that preserve the ground state 2$\times$2$\times$2 CDW structure, and an E$_{g}$-symmetry amplitude mode near 79 cm$^{-1}$.\cite{14,15}  These CDW amplitude modes are associated with the soft zone boundary TA phonon at the L-point,\cite{14,15} which is folded to the zone center when the unit cell is doubled below T$_{\rm CDW}$.\cite{16}

Fig. 2 summarizes the A$_{1g}$ amplitude mode frequency (squares)
and linewidth (circles) in Cu$_{x}$TiSe$_{2}$ as functions of (a) Cu
concentration, x (for T = 6 K) and (b) temperature (for x = 0).  The
A$_{1g}$ amplitude mode frequency and linewidth data were extracted
from Lorentzian fits to the data, as illustrated for some select
spectra in Fig. 1.  Because the A$_{1g}$ amplitude mode (dashed
lines, Fig. 1) is in most cases well-separated from, or much broader
and stronger than, nearby optical modes (dotted lines, Fig. 1),
estimated errors in the amplitude mode frequency obtained in this
manner were $\leq$ 1\%. Fig. 2(b) illustrates that the A$_{1g}$
amplitude mode of Cu$_{x}$TiSe$_{2}$ exhibits temperature-dependent
soft mode behavior typical of amplitude modes observed in other CDW
systems,[17,18] including: (i) a temperature-dependence (filled
squares, Fig. 2(b)) given by the power law form $\omega_{o}$(T)
$\sim$ (1 - T/T$_{\rm CDW}$)$^{\beta}$ with $\beta$ $\sim$ 0.15
(solid line, Fig. 2(b)), (ii) a weakening of the amplitude mode
intensity as the CDW lattice loses coherence as T $\rightarrow$
T$_{CDW}$, and (iii) a dramatic increase in linewidth with
increasing temperature (filled circles, Fig. 2(b)); the latter
mainly reflects overdamping of the amplitude mode due to an increase
in CDW fluctuations, as it is in substantial excess of the
broadening expected from anharmonic (i.e., two-phonon) contributions
(dashed line, Fig. 2(b)).  The anomalous temperature dependence of
the 118 cm$^{-1}$ A$_{1g}$ amplitude mode in Cu$_{x}$TiSe$_{2}$
confirms that its eigenvector couples strongly to the lattice
distortion responsible for the CDW transition at T$_{\rm CDW}$, and
thus mimics the collapse of the soft mode and the CDW gap as T
$\rightarrow$ T$_{\rm CDW}$.

\begin{figure}[tb]
\centerline{\includegraphics[width=8cm]{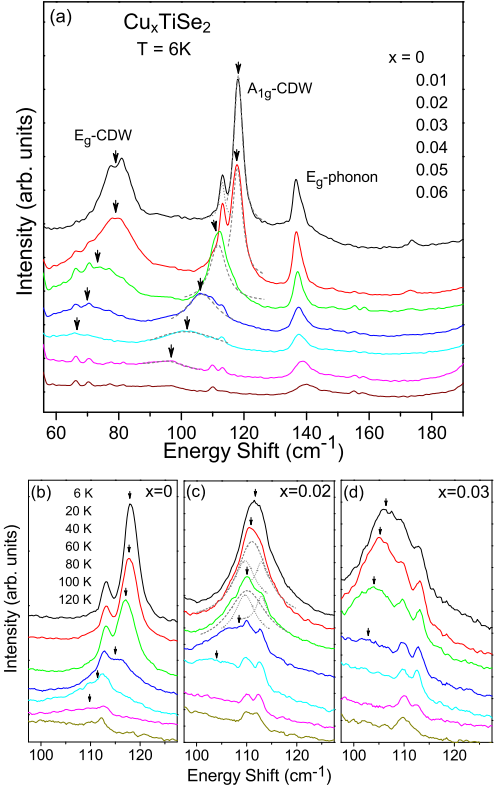}} \caption{(a).
Doping (x) dependence of the Raman spectrum of Cu$_{x}$TiSe$_{2}$ at
T = 6 K, illustrating the E$_{g}$- and A$_{1g}$-symmetry CDW
`amplitude' modes (arrows) as a function of x.  (b)-(d) Temperature
dependence of the A$_{1g}$-symmetry CDW mode spectra at (b) x = 0,
(c) x = 0.02, and (d) x = 0.03 in Cu$_{x}$TiSe$_{2}$.  The spectra
have been offset for clarity, and the vertical scales in (c) and (d)
are respectively 3$\times$ and 4$\times$ the vertical scale in (b).
 Also shown are example Lorentzian fits to the A$_{1g}$ amplitude mode (dashed) and nearby optical modes (dotted).} \label{figure1}
\end{figure}

Significantly, Fig. 1(a) shows that both the 79 cm$^{-1}$ E$_{g}$
and 118 cm$^{-1}$ A$_{1g}$ amplitude modes in Cu$_{x}$TiSe$_{2}$
also exhibit x-dependent mode softening that has nearly identical
characteristics to temperature-dependent mode softening in
Cu$_{x}$TiSe$_{2}$.  For example, Fig. 2(a) shows that the A$_{1g}$
amplitude mode softens by $\sim$18\% between x = 0 and x = 0.05 at T
= 6 K (solid squares), and exhibits a 400\% increase in linewidth
between x = 0 and x = 0.04 at T = 6 K (solid circles).  Note that
the dramatic x-dependent increase in the T = 6 K amplitude mode
linewidths (filled circles, Fig. 2(a)) cannot be attributed to the
effects of disorder (e.g., inhomogeneous broadening caused by Cu
substitution), as there is not a comparably large increase in the
other phonon linewidths with increasing x.  Rather, x-dependent
damping of the CDW amplitude modes in Cu$_{x}$TiSe$_{2}$ reflects a
dramatic enhancement of CDW fluctuations - and a loss of CDW
coherence - with increasing x.  A microscopic analysis of the nature
and origin of x-dependent mode softening in Cu$_{x}$TiSe$_{2}$ (see
Fig. 2(a)) can be made using Rice and collaborators' mean-field
result for the frequency of a CDW amplitude mode,\cite{12}
\begin{equation}
\omega_{o} = 1.4\lambda^{1/2}\tilde{\omega}t^{1/2},
\end{equation}
which has been successfully applied to the analysis of CDW soft mode behavior in other
dichalcogenides.\cite{17} In Eq. 1, $\tilde{\omega}$ is the unscreened (high temperature) phonon frequency, t = (T$_{\rm CDW}$-T)/T$_{\rm CDW}$ is the reduced temperature, and $\lambda$ = N(0)g$^{2}$(0)/$\tilde{\omega}$ is the electron-phonon coupling constant associated with the CDW, where g(0) is the electron-phonon coupling matrix element between the soft mode phonon and the electronic states at the Fermi surface involved in the CDW transition, and N(0) is the joint density of states of the electrons and holes involved in the CDW transition.\cite{17}  Note that the unscreened frequency $\tilde{\omega}$ in Eq. (1) is not expected to have a significant doping dependence between x = 0 and x = 0.06 in Cu$_{x}$TiSe$_{2}$, which is supported by the fact that the optical phonon frequencies exhibit a negligible change with x (e.g., see 137 cm$^{-1}$ mode in Fig. 1).  Consequently, Eq. 1 suggests that the x-dependent softening of the A$_{1g}$ amplitude mode in Cu$_{x}$TiSe$_{2}$ is associated with a substantial reduction in the electron-phonon coupling constant $\lambda$ with doping in Cu$_{x}$TiSe$_{2}$.  One possible source of this reduction is a decrease in the density of electrons/holes participating in the CDW transition between x = 0 and x = 0.05 in Cu$_{x}$TiSe$_{2}$, N(0).  Indeed, Morosan \emph{et al.} observed a $\sim$50\% decrease in the size of the magnetic susceptibility drop below T$_{\rm CDW}$ between x = 0 and x = 0.05 in Cu$_{x}$TiSe$_{2}$, indicating that fewer electronic states are gapped at the CDW transition with increasing x.\cite{5}  Importantly, however, this reduction in N(0) is not likely associated with a loss of Fermi surface nesting with increasing x, because ARPES studies of Cu$_{x}$TiSe$_{2}$ have shown that nesting actually increases with doping.\cite{7}  Another possibility is that the primary effect of Cu intercalation on the CDW phase is caused by the linear expansion of the a-axis parameter with Cu intercalation in Cu$_{x}$TiSe$_{2}$,\cite{5} which leads to a reduction in $\lambda$ by expanding the Ti-Se bond length primarily responsible for the CDW instability in 1$T$-TiSe$_{2}$.\cite{6,16,19}  Notably, this alternative is consistent with Castro Neto's proposal that the layered dichalcogenides have a critical lattice spacing above which CDW order is suppressed.\cite{20}

\begin{figure}[tb]
\centerline{\includegraphics[width=8cm]{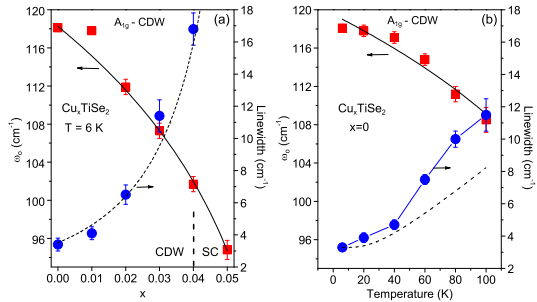}} \caption{(a).
Summary of the T = 6 K A$_{1g}$-symmetry CDW mode frequency
($\omega_{o}$) vs. x (filled squares) and the T = 6 K
A$_{1g}$-symmetry CDW mode linewidth (FWHM) vs. x (filled circles)
for Cu$_{x}$TiSe$_{2}$.  The solid line is a fit to the doping
dependence of the T = 6 K frequency data using
$\omega_{o}$/$\omega_{o}$(0)=(1 - x/x$_{c}$)$^{\beta}$ with $\beta$
$\sim$ 0.15 and x$_{c}$ $\sim$ 0.07, and the dashed line is a fit to
$\Gamma$ $\sim$ (x$_{c}$ - x)$^{-2}$  with x$_{c}$ $\sim$ 0.07. (b).
Summary of the x = 0 A$_{1g}$-symmetry CDW mode frequency
($\omega_{o}$) vs. temperature (filled squares) and the x = 0
A$_{1g}$-symmetry CDW mode linewidth (FWHM) vs. temperature (filled
circles) for TiSe$_{2}$.  The solid line is a fit to the frequency
data with $\omega_{o}$/$\omega_{o}$(0) = (1 - T/T$_{\rm
CDW}$)$^{\beta}$ with $\beta$ $\sim$ 0.15.  The dashed line
illustrates the contribution to the linewidth expected from
two-phonon damping.  The estimated errors from the Lorentzian fits
are $\leq$ 1\% for the frequency data and $\leq$ 8\% for the
linewidth data.  The linewidth data includes a $\sim$ 0.5 cm$^{-1}$
contribution from instrumental broadening that has been accounted
for in fits to the data.} \label{figure2}
\end{figure}

\begin{figure}[tb]
\centerline{\includegraphics[width=8.5cm]{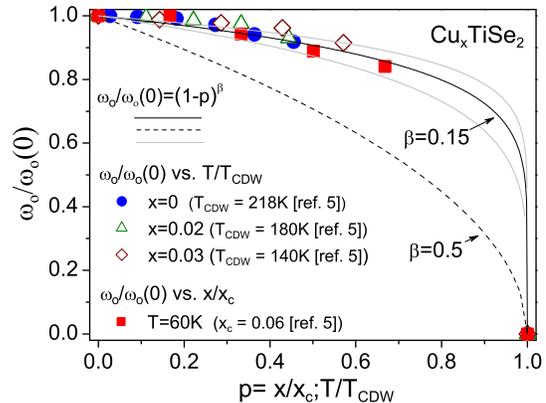}}
\caption{Plots of $\omega_{o}$/$\omega_{o}$(0) vs. x/x$_{c}$ for T =
60 K (x$_{c}$ = 0.06 \cite{5}) (filled squares), and
$\omega_{o}$/$\omega_{o}$(0) vs. T/T$_{\rm CDW}$ for x = 0 (T$_{\rm
CDW}$ = 218 K \cite{5}) (filled circles), x = 0.02 (T$_{\rm CDW}$ =
180 K \cite{5}) (open triangles), and (iv) x = 0.03 (T$_{\rm CDW}$ =
140 K \cite{5}) (open diamonds).  The top dashed line in Fig. 2(c)
shows that all these data sets collapse onto the same curve given by
$\omega_{o}$/$\omega_{o}$(0) = (1 - p)$^{\beta}$, where $\beta$ =
0.15 and p = x/x$_{c}$ or T/T$_{\rm CDW}$.  The gray lines provide
an estimate of the uncertainty in $\beta$ by comparing fits with
$\beta$ = 0.2 (bottom) and $\beta$ = 0.10 (top).  The estimated
errors for the frequency data ($\leq$ 1\%) are roughly given by the
symbol sizes.}
\end{figure}

Remarkably, Fig. 3 shows that, in spite of the different microscopic
effects of doping (x) and temperature on the lattice, x-dependent
and thermal mode softening in Cu$_{x}$TiSe$_{2}$ exhibit essentially
identical scaling behavior.  In particular, Fig. 3 compares the
following data sets:  (i) the normalized A$_{1g}$ amplitude mode
frequency $\omega_{o}$/$\omega_{o}$(0) vs. the reduced doping
x/x$_{c}$ for T = 60 K (using x$_{c}$ = 0.06 from ref. 5) (filled
squares); and (ii) the normalized A$_{1g}$ amplitude mode frequency
$\omega_{o}$/$\omega_{o}$(0) vs. the reduced temperature T/T$_{\rm
CDW}$ for x = 0 (using T$_{\rm CDW}$ = 218 K from ref. 5) (filled
circles), x = 0.02 (T$_{\rm CDW}$=180K \cite{5}) (open triangles),
and x = 0.03 (T$_{\rm CDW}$=140K \cite{5}) (open diamonds) (similar
results are obtained from the E$_g$ amplitude mode frequency). The
solid line in Fig. 3 shows that all these data sets collapse onto
the same curve given by $\omega_{o}$/$\omega_{o}$(0) = (1 -
p)$^{\beta}$, with p = x/x$_{c}$ or T/T$_{\rm CDW}$ and $\beta$
$\sim$ 0.15.  An estimated uncertainty of  $\Delta\beta$ = $\pm$0.05
in the ``best fit'' value of $\beta$ $\sim$ 0.15 is suggested by the
gray lines in Fig. 3, which show the functional form
$\omega_{o}$/$\omega_{o}$(0) = (1 - p)$^{\beta}$  for both $\beta$ =
0.20 (bottom gray line) and $\beta$ = 0.10 (top gray line).

Two key points should be made regarding Fig. 3:  First, although we
cannot measure the amplitude mode frequency with temperature
(doping) all the way to the critical point T$_{CDW}$ (x$_{c}$), for
reasons described above, we can nevertheless extract a reliable
value for the scaling parameter $\beta$ from fits to these data in
Fig. 3 because we know the values of T$_{CDW}$ (for all x) and
x$_{c}$ (at T = 60 K) from ref. 5.  Second, we note that the scaling
parameter $\beta$ $\sim$ 0.15 obtained from our amplitude mode data
is substantially smaller than the value of 1/2 suggested by the
mean-field model of Eq. 1.\cite{12}  However, the scaling parameter
of $\beta$ $\sim$ 0.15 in Cu$_{x}$TiSe$_{2}$ is consistent with the
critical exponent for the order parameter in the 2D three-state
Potts model, $\beta$ = 0.133;\cite{21} this model---which has the
same symmetry as the free energy used by McMillan for the layered
dichalcogenides\cite{14}---is appropriate for Cu$_{x}$TiSe$_{2}$
because of the 3 commensurate CDWs in this material.\cite{6,11}

The `universal' scaling of the Cu$_{x}$TiSe$_{2}$ amplitude mode as
functions of both T/T$_{CDW}$ and x/x$_{c}$ in Fig. 3 emphasizes
that x-dependent mode softening - like more conventional thermal
mode softening - is associated with a critical point at x$_{c}$ that
drives both critical softening behavior ($\omega_{o}$ $\rightarrow$
0) and overdamping of the amplitude mode as x $\rightarrow$ x$_{c}$.
Indeed, the x-dependent amplitude mode softening data between 0 $<$
T $\leq$ 100 K, which are summarized in Fig. 4, suggest the presence
of a CDW phase boundary line x$_{c}$(T) in Cu$_{x}$TiSe$_{2}$ that
extends from the phase boundary line established by Morosan \emph{et
al.}\cite{5} down to a quantum (T$\sim$0) critical point.  To obtain
quantitative estimates of x$_{c}$(T) in Cu$_{x}$TiSe$_{2}$ from our
data, the $\omega_{o}$ vs x curves in Fig. 4 were fit using the same
functional form as that used to fit the x-dependent data at T = 60 K
in Fig. 3, i.e., $\omega_{o}$(x)= $\omega_{o}$(0)(1 -
x/x$_{c}$)$^{\beta}$ with $\beta$ $\sim$ 0.15, where $\omega_{o}$(0)
is the x = 0 value of the A$_{1g}$ amplitude mode frequency at a
particular temperature.  Note that the quantities contributing to
the prefactor, $\omega_{o}$(0), should not have a significant
x-dependence:  the reduced temperature factor in Eq. (1) varies less
than 5\% between x = 0 and x = 0.06, and the unscreened frequency
$\tilde{\omega}$ in Eq. (1) is not expected to have a significant
doping dependence in Cu$_{x}$TiSe$_{2}$ for reasons described above.
Additionally, in obtaining estimates of x$_{c}$(T) from our data, we
assume that the scaling parameter $\beta$ $\sim$ 0.15 obtained from
fits to the T = 60 K data in Fig. 3 doesn't vary significantly for
the $\omega_{o}$ vs x curves at the other temperatures shown in Fig.
4 - this assumption is justified by the wide range of $\omega_{o}$
vs. T and $\omega_{o}$ vs. x curves that scale according the power
law form $\omega_{o}$ $\sim$ (1 - p)$^{0.15}$ in Fig. 3 (p =
T/T$_{\rm CDW}$ or x/x$_{c}$), and by theoretical predictions for
the critical exponent expected for the order parameter in a 2D
system with a three-fold degenerate ground state, $\beta$ =
0.133.\cite{11,21}.  Thus, the resulting fits of the data in Fig. 4
(solid lines) have only x$_{c}$(T) as an unconstrained parameter.
The estimates of T(x$_{c}$) obtained from the fits in Fig. 4 are
represented by the filled circles in the inset of Fig. 4; also shown
for comparison are previous measurements of T(x$_{c}$) (open
circles) by Morosan \emph{et al.}\cite{5}  The reasonableness of our
T(x$_{c}$) estimates is supported by two self-consistency checks:
First, our estimates of x$_{c}$(T) provide good fits of the
x-dependent A$_{1g}$ amplitude mode linewidths using the functional
form, $\Gamma$(T) $\sim$ (x$_{c}$(T) - x)$^{-\gamma}$, as
illustrated by the dashed line in Fig. 2(a) for $\gamma$ $\sim$ 2;
and second, our estimated value for x$_{c}$(T = 80 K) overlaps with
the known phase boundary line from Morosan \emph{et al.}\cite{5}

\begin{figure}[tb]
\centerline{\includegraphics[width=8.5cm]{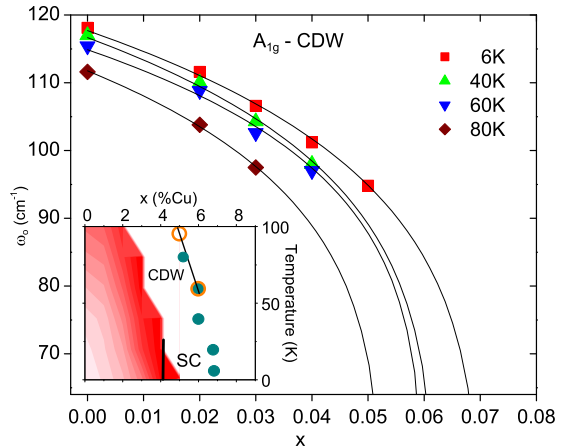}}
\caption{Summary of the A$_{1g}$ amplitude mode frequency
$\omega_{o}$) vs. x for different temperatures.  The solid lines are
fits using $\omega_{o}$/$\omega_{o}$(0) = (1 -
x/x$_{c}$(T))$^{\beta}$, w/ $\beta$ = 0.15-0.16.  (inset)  Estimated
values of T(x$_{c}$) from the fits (filled circles); T$_{\rm CDW}$
data from ref. 5 (open circles) is shown for comparison.  The inset
also shows a contour plot of the temperature and x-dependent
linewidth $\Gamma$ data, which ranges from light red ($\Gamma$
$\sim$ 3 cm$^{-1}$) to dark red ($\Gamma$ $\sim$ 17 cm$^{-1}$).  The
solid line denotes the superconductor (SC) phase boundary line.  The
estimated errors for the frequency data ($\leq$ 1\%) and T(x$_{c}$)
values are roughly given by the symbol sizes.} \label{figure4}
\end{figure}

The values of T(x$_{c}$) estimated from our x-dependent mode
softening results in Fig. 4 are consistent with a low temperature
CDW phase boundary in Cu$_{x}$TiSe$_{2}$ that extends from the phase
boundary line measured by Morosan \emph{et al.}\cite{5} down to a
quantum critical point at roughly x$_{c}$(T = 0) $\sim$ 0.07.  This
suggests that SC and fluctuating CDW order coexist in the doping
range x $\sim$ 0.04 - 0.07 of Cu$_{x}$TiSe$_{2}$.  Furthermore, this
result suggests that the Cu$_{x}$TiSe$_{2}$ phase diagram is
consistent with the T vs. lattice parameter phase diagram plotted by
Castro Neto for the layered dichalcogenides, in which there are two
quantum critical points as a function of increasing lattice
parameter: superconductivity (SC) and CDW order coexist above the
lower of the two critical lattice parameters, while SC is present,
but CDW order is not, above the higher of the two critical lattice
parameters.\cite{20}  We note, however, that because the amplitude
mode becomes overdamped and unobservable very close to the
transition region - due to the breakdown of long-range CDW order and
zone-folding \cite{17,18} - we cannot rule out the possibility that
other effects, e.g., disorder from Cu intercalation, may lead to
different quantum critical behavior (i.e., for T $\sim$ 0 and near x
$\sim$ 0.07) than that implied by the x-dependent scaling behavior
we observe up to x = 0.05 in Fig. 4.  Consequently, it would be
useful to study the putative transition region x$_{c}$(T = 0) $\sim$
0.07 with methods more sensitive to short-range, fluctuating CDW
order, such as inelastic x-ray or neutron scattering.  It is
nevertheless interesting that the value of the quantum critical
point x$_{c}$(T = 0) $\sim$ 0.07 estimated from our x-dependent mode
softening data is close to the peak in T$_{c}$(x), suggesting a
possible connection between SC and the presence of fluctuating CDW
order in Cu$_{x}$TiSe$_{2}$. Indeed, the inset of Fig. 4 also shows
a contour plot of the temperature and x-dependent A$_{1g}$ amplitude
mode linewidth, $\Gamma$, which ranges from light red ($\Gamma$
$\sim$ 3 cm$^{-1}$) to dark red ($\Gamma$ $\sim$ 17 cm$^{-1}$),
illustrating the dramatic increase of CDW fluctuations as the phase
boundary is approached with increasing x and/or temperature. Another
indirect connection between quantum critical behavior and the
expansion of the lattice in Cu$_{x}$TiSe$_{2}$ is also suggested by
the fact that pressure studies of 1$T$-TiSe$_{2}$ \cite{14} don't
show evidence for pressure-induced T$\sim$0 softening of the
A$_{1g}$ amplitude mode that would indicate the presence of a
quantum critical point; this suggests that pressure (lattice
compression) and Cu intercalation (lattice expansion) have
fundamentally different effects on the quantum phases of
1$T$-TiSe$_{2}$.

This material is based on work supported by the U.S. Department of Energy, Division of Materials Sciences, under Award Nos. DE-FG02-07ER46453 and DE-FG02-98-ER45706.  We would like to acknowledge A. Castro Neto and M. V. Klein for useful comments.


\end{document}